\documentclass[aps,twocolumn,amsmath,amssymb,preprintnumbers,superscriptaddress,floatfix]{revtex4}

\usepackage{physics}

\usepackage[utf8]{inputenc}
\usepackage{newtxtext}
\usepackage[upint]{newtxmath}
\usepackage{microtype}
\usepackage{textcomp}
\usepackage{dsfont}
\usepackage{eucal}
\usepackage{siunitx}
\usepackage{soul}

\usepackage{enumerate}
\usepackage{amsfonts}
\usepackage{color}
\usepackage{soul}

\usepackage{todonotes}
\presetkeys%
    {todonotes}%
    {inline}{}

\usepackage{graphicx}

\usepackage[colorlinks,allcolors=blue]{hyperref}
\usepackage[capitalize]{cleveref} 
\usepackage{cleveref}

\newcommand{\me}[1]{\mathrm{e}^{#1}}                            

\renewcommand{\v}[1]{\boldsymbol{#1}}                           
\newcommand{\uv}[1]{\v{e}_{#1}}       

\definecolor{DarkBlue}{rgb}{0,0,0.80}
\definecolor{DarkRed}{rgb}{0.80,0,0}
\definecolor{Purple}{rgb}{0.55,0,0.55}
\definecolor{Purple}{rgb}{0,0,0.8}

\DeclareMathOperator{\diag}{diag}

\newcommand*{\defeq}{\coloneqq}

\newcommand{\up}{\uparrow}                                      
\newcommand{\dn}{\downarrow}                                    

\newcommand{\thouless}{\varepsilon_{\textsc t}}

\newcommand{\eg}{e.g.\ }

\let\epsilon\varepsilon

\begin{document}
\title{Temporarily enhanced superconductivity from magnetic fields}

\author{Eirik Holm Fyhn}
\affiliation{Center for Quantum Spintronics, Department of Physics, Norwegian \\ University of Science and Technology, NO-7491 Trondheim, Norway}
\author{Jacob Linder}
\affiliation{Center for Quantum Spintronics, Department of Physics, Norwegian \\ University of Science and Technology, NO-7491 Trondheim, Norway}

\date{\today}
\begin{abstract}
  \noindent {Contrary to the expected detrimental influence on superconductivity when applying a magnetic field, we predict that the abrupt onset of such a field can temporarily strongly enhance the superconducting order parameter. Specifically, we find that the supercurrent in a Josephson junction with a normal metal weak link can increase more than twentyfold in this way. The effect can be understood from the interplay between the energy-dependence of Andreev reflection and the abrupt spin-dependent shift in the distribution functions for excitations in the system. The duration of the increase depends on the inelastic scattering rate in the system and is estimated to be in the range of nanoseconds. We demonstrate this by developing a method which solves the Usadel equation for an arbitrary time-dependence. This enables the study of ultrafast time-dependent physics in heterostructures combining superconductors with different types of materials.}

\end{abstract}
\maketitle
\textit{Introduction}. Time-dependent phenomena in superconductors encompass a variety of both applied and fundamental physics. These phenomena range from the perfect voltage-to-frequency conversion via the AC Josephson effect to excitation of the amplitude mode of the superconducting order parameter, which is the condensed-matter equivalent of the Higgs boson in the Standard Model.

More recently, interest in time-dependent phenomena in superconductors has been generated by experiments showing optically induced transient states with superconducting properties well above the equilibrium critical temperature~\cite{mitrano2016,nicoletti2014,kaiser2014}.
In superconducting heterostructures it has also been shown that microwaves can greatly increase the critical current~\cite{warlaumont1979,notarys1973}.
This was given a theoretical explanation based on quasiclassical Green's functions~\cite{virtanen2010}.
Another application of quasiclassical Green's functions has been to show that time-dependent exchange fields can produce odd-frequency superconductivity which survives for long distances inside ferromagnets~\cite{houzet2008,bergeret_prl_01}.
This is a type of superconductivity that is interesting due to its non-local temporal symmetry, its direct connection to Majorana states~\cite{linder2019}, and for its resilient nature, making it practically relevant in \eg superconducting spintronics~\cite{linder2019}.

Discovering new time-dependent physical phenomena in superconducting structures, and explaining existing experimental results, is clearly of substantial interest.
Unfortunately, a solution of the quasiclassical Green's function equation is generally not attainable, even numerically, when the system evolves in time.
This is because the relevant equations, presented below, are complicated partial differential equations of infinite order.
So far, approximate solutions have been found for periodic~\cite{cuevas2006,semenov2016,virtanen2010,houzet2008,linder2016} and slow~\cite{tobin1981,kubo2019} temporal evolutions.
Although many situations are either slow or periodic, there is still a multitude of physical systems that are unsolvable with current techniques.
For instance, the transient behaviour of any sudden change that is not periodic, such as a sudden increase in the applied magnetic field or voltage, would not be possible to study, even numerically, with these methods.
Finding a way to solve the Usadel equation that is less restrictive on how it allows the system to evolve in time would therefore open the possibility to study a vast range of new physical phenomena.

Here, we accomplish this goal and present a method solving the time-dependent Usadel equation in hybrid nanostructures that places no constraint on the time-dependence.
We apply this to a superconductor-normal metal-superconductor (SNS) Josephson junction with a time-dependent spin-splitting applied to the N part.
Interestingly, we find that the transient behaviour can involve a large increase in both the supercurrent and the superconducting order parameter.
This is our main result, which stands in stark contrast to the equilibrium effect of an applied magnetic field, which is to exponentially dampen superconductivity~\cite{buzdin2005}. 

In addition to the curious enhancement of superconductivity, which we suggest can be understood as the interplay between properties of Andreev reflection and the transient behaviour of the distribution function, we show how the methodology developed herein can be used to uncover new physics in a wide range of systems. 
It only requires that the proximity effect is sufficiently weak.
In particular, it could be used to study the mostly unexplored territory of explicit time-dependence in odd-frequency superconducting condensates, both in the ballistic and diffusive limit.

\textit{Equations and notation}. 
The quasiclassical theory is valid when the Fermi wavelength is much shorter than all other length scales.
Here we shall focus on the dirty limit, which is valid when the mean free path is short.
However, we note that the same derivation can be done with arbitrary impurity concentration, something that is further discussed in the supplementary material.
The relevant equation for the dirty limit is the Usadel equation~\cite{usadel1970,rammer1986},
\begin{equation}
  D\tilde\nabla\circ \left(\check g \circ \tilde\nabla\circ \check g\right) + i(\check \sigma \circ \check g - \check g\circ \check\sigma) = 0.
  \label{eq:usadel}
\end{equation}
Here, $D$ is the diffusion coefficient, the $8\times 8$ matrix
\begin{equation}
  \check g = \mqty(\hat g^R & \hat g^K \\ 0 & \hat g^A)
\end{equation}
is the isotropic part of the impurity averaged quasiclassical Green's function, 
$\check\sigma$ is a self-energy that depend on the specific system and
\begin{equation}
  \tilde\nabla\circ \check g = \nabla \check g  - ie\left(\hat{\v a} \circ \check g - \check g\circ\hat{\v a}\right)
\end{equation}
is the covariant derivative.
The vector $\hat{\v a}$ includes the effect of the vector potential, but it could also incorporate spin-orbit effects~\cite{bergeret2013,amundsen2017}.
The electron charge is $e = -\abs{e}$.
Finally, the circle-product is 
\begin{equation}
  a\circ b = \exp(\frac i 2 \partial_\varepsilon^a \partial_T^b - \frac i 2 \partial_T^a \partial_\varepsilon^b)ab,
  \label{eq:circProd}
\end{equation}
which is what makes \cref{eq:usadel} difficult when the constituents depend on the center-of-mass time $T$.
The superscripts in \cref{eq:circProd} denote which function the operators acts on and $\varepsilon$ is energy.
The superscripts $R$, $K$ and $A$ will be used to denote the upper left, upper right and lower right $4\times 4$ blocks of $8\times 8$ matrices, respectively.

\Cref{eq:usadel} can be made dimensionless by dividing every term by the Thouless energy, $\thouless \defeq D/L^2$,
where $L$ is { the length of the system.
With this one can define dimensionless quantities, where lengths are given in multiples of $L$ and energies are given in multiples of $\thouless$.
Dimensionless quantities will be used in the rest of this paper.
We also use natural units throughout, meaning that $c = \hbar = 1$.

Quasiclassical theory is invalid at interfaces between different materials.
Consequently, one needs boundary conditions in order to connect the Green's functions in different materials.
Here we use the Kupriyanov-Lukichev boundary condition~\cite{KL1988},
\begin{equation}
  \uv n \cdot \left(\check g_i \circ \tilde\nabla\circ  \check g_i\right) = \frac{z}{2}\left(\check g_i\circ \check g_j - \check g_j\circ \check g_i\right),
  \label{eq:KL}
\end{equation}
which is valid for low-transparency tunneling interfaces.
The subscripts $i$ and $j$ labels the two different regions, the unit normal vector $\uv n$ points out of region $i$ and $z$ is the ratio between the bulk resistance of a part of the material that is of length $L$ and the interface resistance.
Although we use the Kupriyanov-Lukichev boundary condition here, the same method could also be used with other types of boundaries~\cite{eschrig2015}.

The quasiclassical Green's function satisfies the normalization condition
$\check g \circ \check g = 1$
and the relations
\begin{align}
\label{eq:rels}
  \hat g^A = -\hat \rho_3\left(\hat g^R\right)^\dagger\hat \rho_3,
  \qquad
  \hat g^K = \hat g^R\circ h - h\circ \hat g^A,
\end{align}
where $\hat \rho_3 = \diag(1,1,-1,-1)$.
From \cref{eq:rels} one can see that it is sufficient to solve for the retarded Greens function $\hat g^R$ and the distribution function $h$.
\Cref{eq:usadel} does not fully specify $h$, and we can use this freedom to make $h$ block-diagonal~\cite{schmid1975}.

Finally, we use capital letters to denote Fourier transforms,
\begin{equation}
  F(t, T, \v r) \equiv \mathcal{F}(f)(t, T, \v r) = \frac 1 {2\pi}\int_{-\infty}^\infty \dd{\varepsilon} f(\varepsilon, T, \v r) \me{-i \varepsilon t}.
  \label{eq:FT}
\end{equation}
and $\bullet$ to denote the circle-product between functions of the relative time $t$.
That is, $\bullet$ is the mathematical operation which satisfies $\mathcal{F}(f\circ g) = F\bullet G$.

The aim is to find the Green's function that solves \cref{eq:usadel} in a region that is connected through the boundary condition in \cref{eq:KL} to a region with Green's function $\check g_s$.
This region could for instance be a superconducting reservoir.
We have developed a method which solves the Usadel equation with an arbitrary time-dependence, allowing for the study of quantum quenches and ultrafast dynamics, and present this method below.

The first step is to write the retarded Green's function as $\hat g^R = \hat \rho_3 + \hat g + \hat f$, where $\hat g$ and $\hat f$ are block-diagonal and block-antidiagonal, respectively.
Under the assumption that the proximity effect is small, the components of $\hat g$ and $\hat f$ are all much smaller than one.
One way to formalise this is to Taylor expand $\hat g$ and $\hat f$ in terms of the interface parameter $z$.
When $\check\sigma^R$ is block-diagonal and $z=0$, we find that $\hat g^R = \hat \rho_3$ solves the Usadel equation.
Hence, assuming $\check \sigma^R$ is block-diagonal to lowest order in $z$, we can write
\begin{equation}
  \hat f = \sum_{n=1}^\infty z^n \hat f_n \quad \text{and} \quad \hat g = \sum_{n=1}^\infty z^n \hat g_n.
\end{equation}
From the normalization condition $\hat g^R \circ \hat g^R = 1$, we see that
$2\hat\rho_3 \hat g + \hat g\circ \hat g =-\hat f\circ \hat f$ and
$\hat g\circ \hat f = -\hat f\circ \hat g$.
Hence, $\hat g_1 = 0$ and $\hat g_2 = -\frac 1 2 \hat\rho_3\hat f_1\circ \hat f_1$.

To first order in $z$, the retarded part of the Usadel equation reads
\begin{equation}
  \hat\rho_3\tilde\nabla\circ\left(\tilde\nabla\circ \hat f_1\right)
  + 2i\varepsilon\hat\rho_3\hat f_1 
  + i(\hat \sigma^R \circ \hat f_1 - \hat f_1\circ \hat\sigma^R)= 0,
  \label{eq:usadelLinearized}
\end{equation}
where $\varepsilon\hat\rho_3$ has been extracted from the self-energy and $\hat\sigma^R$ is the remaining part.
The self-energy $\hat\sigma^R$ could also depend on $\hat g^R$, for instance if the system included spin-orbit impurity scattering or spin-flip scattering~\cite{linder2008}.
In that case \cref{eq:usadelLinearized} would look slightly different, but the derivation would be similar.
To first order in $z$, the boundary condition~\eqref{eq:KL} reads
\begin{equation}
  \uv n\cdot\tilde\nabla\circ \hat f_1 = \hat f_s.
  \label{eq:KLLinearized}
\end{equation}

Despite being linearized, \cref{eq:usadelLinearized,eq:KLLinearized} are not much simpler than the original Usadel equation and Kupriyanov-Lukichev boundary condition.
They still include the circle-product, given in \cref{eq:circProd}, meaning that they are still PDEs of infinite order.
However, one observation can be made which will drastically simplify the equations.
This is the fact that all the circle-products are between $\hat f_1$ and functions that are independent of energy $\varepsilon$.
It is this fact, not that the equations are linear, that is crucial for the solvability of \cref{eq:usadelLinearized,eq:KLLinearized}.
As we shall see, this observation allows us to evaluate all the circle-products if we first Fourier transform the equations.

When a function $(\varepsilon, T) \mapsto a(T)$ is independent of $\varepsilon$, the Fourier transform, as given by \cref{eq:FT}, is simply
$A(t, T) = \delta(t)a(T)$,
where $\delta$ is the Dirac delta distribution.
Accordingly, the circle products of a function $(\varepsilon, T) \mapsto f(\varepsilon, T)$ with a function $(\varepsilon, T) \mapsto a(T)$ are, in Fourier space,
\begin{subequations}
  \label{eq:circProdSimp}
 \begin{align}
   (A\bullet F)(t, T) = a(T+t/2)F(t, T), \\
  (F\bullet A)(t, T) = F(t, T)a(T-t/2).
 \end{align} 
\end{subequations}
With this, all the circle-products in \cref{eq:usadelLinearized} turn into normal matrix multiplications when evaluated in Fourier-space.
This is under the assumption that the self-energy $\hat\sigma^R$ does not depend explicitly on $\varepsilon$.
However, it can depend implicitly on energy through its dependence on $\check g$, as mentioned above.

Let the subscripts $+$ and $-$ denote
$B_\pm (t, T) = b(T\pm t/2)$.
Then the equations for the retarded Green's function become
\begin{subequations}
\begin{align}
\begin{split}
  2\frac{\partial \hat F_1}{\partial t} 
  &= \begin{aligned}[t]
    &\nabla^2 \hat F_1 + 2i\left(\nabla\hat F_1\cdot \hat{\v A}_- - \hat{\v A}_+\cdot\nabla \hat F_1 \right) \\
    &+ i\left(\hat F_1 \nabla\cdot\hat{\v A}_- - \nabla\cdot\hat{\v A}_+\hat F_1\right)
    -\hat{\v A}_+^2 \hat F_1 + \hat{\v A}_+ \hat F_1\hat{\v A}_- \\
    &- \hat F_1\hat{\v A}_-^2
    + i\hat \rho_3\left(\hat\Sigma^R_+\hat F_1 - \hat F_1\hat\Sigma^R_-\right),
  \end{aligned}
\end{split}
\label{eq:UsadelFinal}
\\
  &\uv n\cdot \left[\nabla\hat F_1 - i\left(\hat{\v A}_+ \hat F_1 - \hat F_1\hat{\v A}_-\right)\right] = \hat F_s.
  \label{eq:KLFinal}
\end{align}
\end{subequations}
Hence, an approximate solution to the full time-dependent Usadel equation can be found by solving a normal PDE of matrices.
The approximation is good as long as the proximity effect is weak and, crucially, no assumptions have been made with regards to the time-dependence.
This approach therefore works for systems that vary both fast and slow in time and regardless of whether or not the system is periodic. 
The equations for the distribution function $h$ can be obtained in a similar way.
This is shown in the supplementary.

\textit{Application.} We now use the above framework to show the counterintuitive result that the abrupt onset of a magnetic field can temporarily strongly increase superconducting order.
Consider an SNS-junction with no vector potential and a time-dependent, spatially uniform exchange field $m(T)$ that lifts the spin-degeneracy of the bands.
The geometry is shown in the inset of \cref{fig:curr} where the nanowire geometry allows us to neglect the orbital effect of the magnetic field whereas the thick superconducting regions screen the effect of the magnetic field in the bulk.
The self-energy associated with the exchange field is $\hat\sigma^R = m \diag(1,-1,1,-1)$.
We also include the effect of inelastic scattering through the relaxation time approximation~\cite{virtanen2010}, which adds
\begin{equation}
  \check\sigma_i = \mqty( i\delta\hat\rho_3  & 2i\delta\hat\rho_3 h_\text{eq} \\0  & -i\delta\hat\rho_3),
\end{equation}
to the self energy. Here $\delta$ is the inelastic scattering rate and $h_\text{eq}(\varepsilon) = \tanh(\beta\varepsilon/2)$, where $\beta$ is the inverse temperature towards which the system relaxes.

If we write the upper right block of $\hat F_1$ as $F_1 = \sigma_1F_t + \sigma_2 F_s$, where $\sigma_1$ and $\sigma_2$ are Pauli matrices, the zeroth order distribution function $H_0 = H_L I_4 + H_{TS}\diag(1,-1,-1,1)$ and let $m^{\pm}(t,T) \defeq m(T+t/2) \pm m(T-t/2)$, we find that
\begin{subequations}
  \label{eq:exchange}
  \begin{align}
  \left(2\frac{\partial}{\partial t} - \nabla^2 + 2\delta\right)\mqty(F_s \\ F_t)
  = \mqty(-m^+ F_t \\ m^+ F_s), \\
  \left.\uv n \cdot \nabla F_s \right\rvert_{x=0,1} = F^\text{BSC}_{l,r}, \quad 
  \left.\uv n \cdot \nabla F_t \right\rvert_{x=0,1} = 0, \\
  \left(\frac{\partial}{\partial T} + 2\delta\right)\mqty(H_L-H_\text{eq} \\ H_{TS})
  = \mqty(-m^- H_{TS} \\ m^- H_L), 
  \end{align}
\end{subequations}
where $F^\text{BCS}_l = \Delta\me{-\delta t} J_0(\abs{\Delta}t)\theta(t)$ and $F^\text{BCS}_r = \me{i\phi}F^\text{BCS}_l$ are the anomalous Green's functions in the left and right superconductors, respectively. $J_0$ is the zeroth order Bessel function of the first kind, $\Delta$ is the superconducting gap parameter and $\phi$ is the phase difference between the two superconductors.
\Cref{eq:exchange} can be solved analytically for arbitrary $m(T)$, and the solution is shown in the supplementary. The interface parameter $z$ is assumed small enough to fulfill the criterion of a weak proximity effect for all relevant times $t$ and $T$.

\begin{figure}[tbp]
  \centering
  \includegraphics[width=1.0\linewidth]{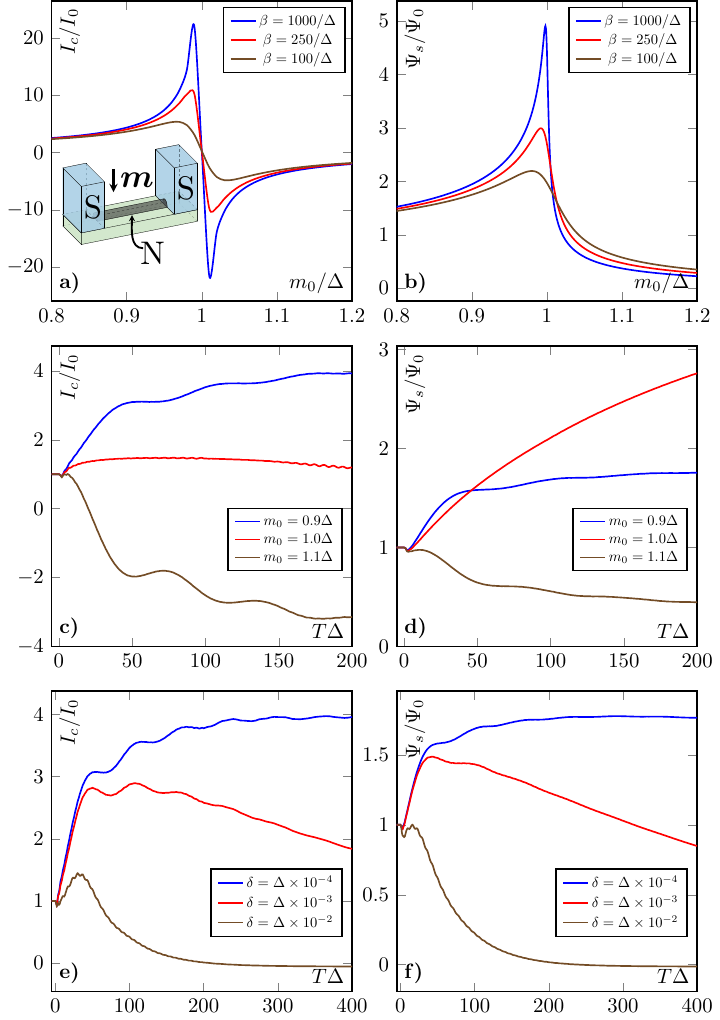}
  \caption{Critical current $I_c$ and singlet Cooper pair correlation function $\Psi_s$ normalized by the values at zero exchange field, $I_0$ and $\Psi_0$. The inset in a) shows a sketch of the setup.
  Panels a) and b) show the steady state values obtained with $\delta = 0$ for various values of exchange field $m_0$.
  Panels c) and d) show the time evolution for different values of $m_0$ with $\delta/\Delta = \num{e-4}$ and $\beta\Delta = 1000$.
  Panels e) and f) show the time evolution for different values of $\delta$ with $m_0/\Delta = 0.9$ and $\beta\Delta = 1000$.
  In all cases $\Delta/\thouless = 10$.
  $\Psi_s$ is computed for $\phi=0$ which makes the enhancement predicted here applicable also to an SN junction.
  }%
  \label{fig:curr}
\end{figure}

Consider an exchange field that abruptly changes value from $0$ to $m_0$ at time $T=0$, $m(T) = m_0 \theta(T)$.
The critical supercurrent
\begin{equation}
  I_c = \max_{\phi\in(0,2\pi]}\frac{\pi N_0 e D}{4}\Tr\left[\hat\rho_3\left(\check G \bullet \tilde\nabla\bullet \check G\right)^K\right]_{t=0},
\end{equation}
 and singlet Cooper pair correlation function 
\begin{equation}
  \Psi_s  = \left.-i\pi N_0\left(F_s\bullet H_L - F_t\bullet H_{TS}\right)\right\rvert_{t=0},
\end{equation}
following an abrupt change in the exchange field are shown in \cref{fig:curr}.
When the time becomes comparable to the inelastic scattering time, both $I_c$ and $\Psi_s$ are suppressed and the quantities reach their equilibrium values.
However, before that, $I_c$ and $\Psi_s$ are significantly enhanced when the exchange field is close to the superconducting gap $\Delta$.
When $m_0 \ll\Delta$ there is only a slight change to the current and Cooper pair correlation function.

We suggest that the behaviour of $I_c$ and $\Psi_s$ can be understood from the interplay between the spin-dependence of the non-equilibrium distribution function and the energy dependence of both the Andreev reflection probability and the degree of coherence between the participating electrons and holes.
To see this, we note that in a time-independent situation both $I_c$ and $\Psi_s$, jointly denoted $A$ below, can be written as an integral over energy of the form
\begin{equation}
  A = \int \dd{\varepsilon}(a_\up h_\up + a_\dn h_\dn),
\end{equation}
as shown in the supplementary. Here, $h_{\up}$ and $h_\dn$ are the distribution functions for electrons with spin $\up$ and $\dn$.
The explanation can be summarized as three key points.

First, $a_{\up,\dn}$ is of largest amplitude at energies close to $\pm\Delta$ and $\pm m$, where $a_\up$ is large close to $\varepsilon=-m$ and $a_\dn$ is large close to $\varepsilon=m$.
These energies are special in the context of Andreev reflections, which is the process relevant for transferring superconductivity into the normal metal.
At $\varepsilon = \pm\Delta$ there is a large peak in the Andreev reflection probability~\cite{BTK1982} which physically can be understood as resonant scattering produced by subsequent reflections by the interface and the superconducting order parameter ~\cite{asano2004}.
At $\varepsilon = \mp m$ the wavevector of the incoming electrons, $k^e_{\up\dn} = \sqrt{2m(E_F + \varepsilon \pm m)}$, match that of the retroreflected holes, $k^h_{\dn\up} = \sqrt{2m(E_F - \varepsilon\mp m)}$.
Hence, at energies close to $\pm m$ the superconducting correlations penetrate far into the normal metal.

Second, $a_{\up,\dn}$ is antisymmetric close to $\mp m$, as long as $m < \Delta$.
This is shown in the supplementary.
That is, filled states with energy just above $\mp m$ contribute oppositely to filled states with energy just below $\mp m$.
Hence, when $m > 0$ and the system is at equilibrium, such that $h_\up(\varepsilon) = h_\dn(\varepsilon) = \tanh(\beta\varepsilon/2)$, the contributions to $\Psi_s$ and $I_c$ are suppressed because the coherent states are shiftet away from the Fermi surface.
However, before inelastic scattering relaxes the system, we find that the distribution functions evolve toward $h_{\up,\dn}(\varepsilon) = \tanh[(\varepsilon \pm m)/\beta]$.
This is physically reasonable since an abrupt temporal change induced by the magnetic field not only shifts the energy levels, but also preserves the occupation of these states before they have had time to relax.
The energy shift in the antisymmetric contribution to $a_{\up,\dn}$ coming from the coherent Andreev reflections are thus matched by a similar shift in the distribution function, so $\Psi_s$ and $I_c$ are not suppressed as $m$ is increased.

Third, when $m\approx \Delta$ the enhanced probability of Andreev reflections amplify the contribution from $\varepsilon \approx m$.
In equilibrium, both the positive and negative contributions are amplified, so the overall effect is still a suppression of $\Psi_s$ and $I_c$ when compared to $m=0$.
However, in the transient period with $h_{\up,\dn}(\varepsilon) = \tanh[(\varepsilon \pm m)/\beta]$ the consequence is a manifold increase in $\Psi_s$ and $I_c$.
In other words, when $m\approx\Delta$ the Andreev reflections with the longest lifetimes are also the ones with the highest probability of occurring, and the non-equilibrium distribution functions that are present before the system has had time to relax allows this to manifest as a strong enhancement in superconductivity.

We find that the time-scale for which the $I_c$ and $\Psi_s$ are able to reach their amplified states is given primarily by $\Delta$.
Hence, in order to experimentally detect the enhanced supercurrent it is necessary that $\delta/\Delta$ is not too large.
From \cref{fig:curr} one can see that $\delta < 10^{-2}\Delta$ is sufficient to observe an increase in the supercurrent.
Experimental values of the inelastic scattering rate, or Dynes parameter, are often found by parameter fitting and values as low as $\delta/\Delta = \num{2.2e-5}$ has been reported in the millikelvin regime~\cite{feschenko2015}.
With $\Delta \approx \SI{1}{\milli\electronvolt}$ and $\delta/\Delta = \num{2.2e-5}$, the relaxation time is about $\SI{10}{\nano\second}$.
A Zeeman splitting of $\SI{1}{\milli\electronvolt}$ is achieved with a magnetic field strength of around $\SI{30}{\tesla}/g$, where $g$ is the Landau factor.
This could be either tens of $\si{T}$ if $g = 2$ or tens of \si{\milli\tesla} when $g\approx \num{e3}$.
The latter can be found for instance in Dirac semimetals~\cite{li2019}.
In the former case, an Ising type superconductor such as NbSe$_2$ can be used to retain superconductivity at high in-plane fields.

The strong enhancement of the proximity-induced singlet order parameter $\Psi_s$ suggests that the order parameter in the superconductor, if solved for self-consistently, could potentially also be enhanced by virtue of the inverse proximity effect.
In turn, this would imply an increase in the critical temperature $T_c$ of the superconducting transition.
We leave this issue, which requires complicated time-dependent, self-consistent numerical calculations, for a future work.

\textit{Conclusion.} We have presented a method for solving the time-dependent Usadel equation with arbitrary time-dependence.
This is made possible by two observations.
First, the circle-products simplifies considerably in Fourier space when one of the arguments are independent of energy.
Second, by linearizing the equations, only such products remain.

We applied this method to analytically study SNS-junction with time-dependent Zeeman-splitting $m$ where a magnetic field is abruptly turned on.
We demonstrated a strong enhancement of the supercurrent and Cooper pair correlation function when $m \approx \Delta$, where $\Delta$ is the superconducting gap.
In particular, if the inelastic scattering rate $\delta$ is smaller than $\Delta\times\num{e-2}$ and the magnetic field changes value during a time frame shorter than $1/\delta$, our results show up to a twentyfold increase in the magnetic field that potentially lasts for tens of nanoseconds.

\begin{acknowledgments}
This work was supported by the Research Council of Norway through grant 240806, and its Centres of Excellence funding scheme grant 262633 ``\emph{QuSpin}''. J. L. also acknowledge funding from the NV-faculty at the Norwegian University of Science and Technology. 
\end{acknowledgments}

\clearpage
\bibliography{bibliography}

\begin{thebibliography}{27}%
\makeatletter
\providecommand \@ifxundefined [1]{%
 \@ifx{#1\undefined}
}%
\providecommand \@ifnum [1]{%
 \ifnum #1\expandafter \@firstoftwo
 \else \expandafter \@secondoftwo
 \fi
}%
\providecommand \@ifx [1]{%
 \ifx #1\expandafter \@firstoftwo
 \else \expandafter \@secondoftwo
 \fi
}%
\providecommand \natexlab [1]{#1}%
\providecommand \enquote  [1]{``#1''}%
\providecommand \bibnamefont  [1]{#1}%
\providecommand \bibfnamefont [1]{#1}%
\providecommand \citenamefont [1]{#1}%
\providecommand \href@noop [0]{\@secondoftwo}%
\providecommand \href [0]{\begingroup \@sanitize@url \@href}%
\providecommand \@href[1]{\@@startlink{#1}\@@href}%
\providecommand \@@href[1]{\endgroup#1\@@endlink}%
\providecommand \@sanitize@url [0]{\catcode `\\12\catcode `\$12\catcode
  `\&12\catcode `\#12\catcode `\^12\catcode `\_12\catcode `\%12\relax}%
\providecommand \@@startlink[1]{}%
\providecommand \@@endlink[0]{}%
\providecommand \url  [0]{\begingroup\@sanitize@url \@url }%
\providecommand \@url [1]{\endgroup\@href {#1}{\urlprefix }}%
\providecommand \urlprefix  [0]{URL }%
\providecommand \Eprint [0]{\href }%
\providecommand \doibase [0]{http://dx.doi.org/}%
\providecommand \selectlanguage [0]{\@gobble}%
\providecommand \bibinfo  [0]{\@secondoftwo}%
\providecommand \bibfield  [0]{\@secondoftwo}%
\providecommand \translation [1]{[#1]}%
\providecommand \BibitemOpen [0]{}%
\providecommand \bibitemStop [0]{}%
\providecommand \bibitemNoStop [0]{.\EOS\space}%
\providecommand \EOS [0]{\spacefactor3000\relax}%
\providecommand \BibitemShut  [1]{\csname bibitem#1\endcsname}%
\let\auto@bib@innerbib\@empty
\bibitem [{\citenamefont {Mitrano}\ \emph {et~al.}(2016)\citenamefont
  {Mitrano}, \citenamefont {Cantaluppi}, \citenamefont {Nicoletti},
  \citenamefont {Kaiser}, \citenamefont {Perucchi}, \citenamefont {Lupi},
  \citenamefont {Di~Pietro}, \citenamefont {Pontiroli}, \citenamefont
  {Ricc{\`o}}, \citenamefont {Clark} \emph {et~al.}}]{mitrano2016}%
  \BibitemOpen
  \bibfield  {author} {\bibinfo {author} {\bibfnamefont {M.}~\bibnamefont
  {Mitrano}}, \bibinfo {author} {\bibfnamefont {A.}~\bibnamefont {Cantaluppi}},
  \bibinfo {author} {\bibfnamefont {D.}~\bibnamefont {Nicoletti}}, \bibinfo
  {author} {\bibfnamefont {S.}~\bibnamefont {Kaiser}}, \bibinfo {author}
  {\bibfnamefont {A.}~\bibnamefont {Perucchi}}, \bibinfo {author}
  {\bibfnamefont {S.}~\bibnamefont {Lupi}}, \bibinfo {author} {\bibfnamefont
  {P.}~\bibnamefont {Di~Pietro}}, \bibinfo {author} {\bibfnamefont
  {D.}~\bibnamefont {Pontiroli}}, \bibinfo {author} {\bibfnamefont
  {M.}~\bibnamefont {Ricc{\`o}}}, \bibinfo {author} {\bibfnamefont {S.~R.}\
  \bibnamefont {Clark}},  \emph {et~al.},\ }\href@noop {} {\bibfield  {journal}
  {\bibinfo  {journal} {Nature}\ }\textbf {\bibinfo {volume} {530}},\ \bibinfo
  {pages} {461} (\bibinfo {year} {2016})}\BibitemShut {NoStop}%
\bibitem [{\citenamefont {Nicoletti}\ \emph {et~al.}(2014)\citenamefont
  {Nicoletti}, \citenamefont {Casandruc}, \citenamefont {Laplace},
  \citenamefont {Khanna}, \citenamefont {Hunt}, \citenamefont {Kaiser},
  \citenamefont {Dhesi}, \citenamefont {Gu}, \citenamefont {Hill},\ and\
  \citenamefont {Cavalleri}}]{nicoletti2014}%
  \BibitemOpen
  \bibfield  {author} {\bibinfo {author} {\bibfnamefont {D.}~\bibnamefont
  {Nicoletti}}, \bibinfo {author} {\bibfnamefont {E.}~\bibnamefont
  {Casandruc}}, \bibinfo {author} {\bibfnamefont {Y.}~\bibnamefont {Laplace}},
  \bibinfo {author} {\bibfnamefont {V.}~\bibnamefont {Khanna}}, \bibinfo
  {author} {\bibfnamefont {C.~R.}\ \bibnamefont {Hunt}}, \bibinfo {author}
  {\bibfnamefont {S.}~\bibnamefont {Kaiser}}, \bibinfo {author} {\bibfnamefont
  {S.~S.}\ \bibnamefont {Dhesi}}, \bibinfo {author} {\bibfnamefont {G.~D.}\
  \bibnamefont {Gu}}, \bibinfo {author} {\bibfnamefont {J.~P.}\ \bibnamefont
  {Hill}}, \ and\ \bibinfo {author} {\bibfnamefont {A.}~\bibnamefont
  {Cavalleri}},\ }\href {\doibase 10.1103/PhysRevB.90.100503} {\bibfield
  {journal} {\bibinfo  {journal} {Phys. Rev. B}\ }\textbf {\bibinfo {volume}
  {90}},\ \bibinfo {pages} {100503(R)} (\bibinfo {year} {2014})}\BibitemShut
  {NoStop}%
\bibitem [{\citenamefont {Kaiser}\ \emph {et~al.}(2014)\citenamefont {Kaiser},
  \citenamefont {Hunt}, \citenamefont {Nicoletti}, \citenamefont {Hu},
  \citenamefont {Gierz}, \citenamefont {Liu}, \citenamefont {Le~Tacon},
  \citenamefont {Loew}, \citenamefont {Haug}, \citenamefont {Keimer},\ and\
  \citenamefont {Cavalleri}}]{kaiser2014}%
  \BibitemOpen
  \bibfield  {author} {\bibinfo {author} {\bibfnamefont {S.}~\bibnamefont
  {Kaiser}}, \bibinfo {author} {\bibfnamefont {C.~R.}\ \bibnamefont {Hunt}},
  \bibinfo {author} {\bibfnamefont {D.}~\bibnamefont {Nicoletti}}, \bibinfo
  {author} {\bibfnamefont {W.}~\bibnamefont {Hu}}, \bibinfo {author}
  {\bibfnamefont {I.}~\bibnamefont {Gierz}}, \bibinfo {author} {\bibfnamefont
  {H.~Y.}\ \bibnamefont {Liu}}, \bibinfo {author} {\bibfnamefont
  {M.}~\bibnamefont {Le~Tacon}}, \bibinfo {author} {\bibfnamefont
  {T.}~\bibnamefont {Loew}}, \bibinfo {author} {\bibfnamefont {D.}~\bibnamefont
  {Haug}}, \bibinfo {author} {\bibfnamefont {B.}~\bibnamefont {Keimer}}, \ and\
  \bibinfo {author} {\bibfnamefont {A.}~\bibnamefont {Cavalleri}},\ }\href
  {\doibase 10.1103/PhysRevB.89.184516} {\bibfield  {journal} {\bibinfo
  {journal} {Phys. Rev. B}\ }\textbf {\bibinfo {volume} {89}},\ \bibinfo
  {pages} {184516} (\bibinfo {year} {2014})}\BibitemShut {NoStop}%
\bibitem [{\citenamefont {Warlaumont}\ \emph {et~al.}(1979)\citenamefont
  {Warlaumont}, \citenamefont {Brown}, \citenamefont {Foxe},\ and\
  \citenamefont {Buhrman}}]{warlaumont1979}%
  \BibitemOpen
  \bibfield  {author} {\bibinfo {author} {\bibfnamefont {J.~M.}\ \bibnamefont
  {Warlaumont}}, \bibinfo {author} {\bibfnamefont {J.~C.}\ \bibnamefont
  {Brown}}, \bibinfo {author} {\bibfnamefont {T.}~\bibnamefont {Foxe}}, \ and\
  \bibinfo {author} {\bibfnamefont {R.~A.}\ \bibnamefont {Buhrman}},\ }\href
  {\doibase 10.1103/PhysRevLett.43.169} {\bibfield  {journal} {\bibinfo
  {journal} {Phys. Rev. Lett.}\ }\textbf {\bibinfo {volume} {43}},\ \bibinfo
  {pages} {169} (\bibinfo {year} {1979})}\BibitemShut {NoStop}%
\bibitem [{\citenamefont {Notarys}\ \emph {et~al.}(1973)\citenamefont
  {Notarys}, \citenamefont {Yu},\ and\ \citenamefont
  {Mercereau}}]{notarys1973}%
  \BibitemOpen
  \bibfield  {author} {\bibinfo {author} {\bibfnamefont {H.~A.}\ \bibnamefont
  {Notarys}}, \bibinfo {author} {\bibfnamefont {M.~L.}\ \bibnamefont {Yu}}, \
  and\ \bibinfo {author} {\bibfnamefont {J.~E.}\ \bibnamefont {Mercereau}},\
  }\href {\doibase 10.1103/PhysRevLett.30.743} {\bibfield  {journal} {\bibinfo
  {journal} {Phys. Rev. Lett.}\ }\textbf {\bibinfo {volume} {30}},\ \bibinfo
  {pages} {743} (\bibinfo {year} {1973})}\BibitemShut {NoStop}%
\bibitem [{\citenamefont {Virtanen}\ \emph {et~al.}(2010)\citenamefont
  {Virtanen}, \citenamefont {Heikkil\"a}, \citenamefont {Bergeret},\ and\
  \citenamefont {Cuevas}}]{virtanen2010}%
  \BibitemOpen
  \bibfield  {author} {\bibinfo {author} {\bibfnamefont {P.}~\bibnamefont
  {Virtanen}}, \bibinfo {author} {\bibfnamefont {T.~T.}\ \bibnamefont
  {Heikkil\"a}}, \bibinfo {author} {\bibfnamefont {F.~S.}\ \bibnamefont
  {Bergeret}}, \ and\ \bibinfo {author} {\bibfnamefont {J.~C.}\ \bibnamefont
  {Cuevas}},\ }\href {\doibase 10.1103/PhysRevLett.104.247003} {\bibfield
  {journal} {\bibinfo  {journal} {Phys. Rev. Lett.}\ }\textbf {\bibinfo
  {volume} {104}},\ \bibinfo {pages} {247003} (\bibinfo {year}
  {2010})}\BibitemShut {NoStop}%
\bibitem [{\citenamefont {Houzet}(2008)}]{houzet2008}%
  \BibitemOpen
  \bibfield  {author} {\bibinfo {author} {\bibfnamefont {M.}~\bibnamefont
  {Houzet}},\ }\href {\doibase 10.1103/PhysRevLett.101.057009} {\bibfield
  {journal} {\bibinfo  {journal} {Phys. Rev. Lett.}\ }\textbf {\bibinfo
  {volume} {101}},\ \bibinfo {pages} {057009} (\bibinfo {year}
  {2008})}\BibitemShut {NoStop}%
\bibitem [{\citenamefont {Bergeret}\ \emph {et~al.}(2001)\citenamefont
  {Bergeret}, \citenamefont {Volkov},\ and\ \citenamefont
  {Efetov}}]{bergeret_prl_01}%
  \BibitemOpen
  \bibfield  {author} {\bibinfo {author} {\bibfnamefont {F.~S.}\ \bibnamefont
  {Bergeret}}, \bibinfo {author} {\bibfnamefont {A.~F.}\ \bibnamefont
  {Volkov}}, \ and\ \bibinfo {author} {\bibfnamefont {K.~B.}\ \bibnamefont
  {Efetov}},\ }\href {\doibase 10.1103/PhysRevLett.86.4096} {\bibfield
  {journal} {\bibinfo  {journal} {Phys. Rev. Lett.}\ }\textbf {\bibinfo
  {volume} {86}},\ \bibinfo {pages} {4096} (\bibinfo {year}
  {2001})}\BibitemShut {NoStop}%
\bibitem [{\citenamefont {Linder}\ and\ \citenamefont
  {Balatsky}(2019)}]{linder2019}%
  \BibitemOpen
  \bibfield  {author} {\bibinfo {author} {\bibfnamefont {J.}~\bibnamefont
  {Linder}}\ and\ \bibinfo {author} {\bibfnamefont {A.~V.}\ \bibnamefont
  {Balatsky}},\ }\href {\doibase 10.1103/RevModPhys.91.045005} {\bibfield
  {journal} {\bibinfo  {journal} {Rev. Mod. Phys.}\ }\textbf {\bibinfo {volume}
  {91}},\ \bibinfo {pages} {045005} (\bibinfo {year} {2019})}\BibitemShut
  {NoStop}%
\bibitem [{\citenamefont {Cuevas}\ \emph {et~al.}(2006)\citenamefont {Cuevas},
  \citenamefont {Hammer}, \citenamefont {Kopu}, \citenamefont {Viljas},\ and\
  \citenamefont {Eschrig}}]{cuevas2006}%
  \BibitemOpen
  \bibfield  {author} {\bibinfo {author} {\bibfnamefont {J.~C.}\ \bibnamefont
  {Cuevas}}, \bibinfo {author} {\bibfnamefont {J.}~\bibnamefont {Hammer}},
  \bibinfo {author} {\bibfnamefont {J.}~\bibnamefont {Kopu}}, \bibinfo {author}
  {\bibfnamefont {J.~K.}\ \bibnamefont {Viljas}}, \ and\ \bibinfo {author}
  {\bibfnamefont {M.}~\bibnamefont {Eschrig}},\ }\href {\doibase
  10.1103/PhysRevB.73.184505} {\bibfield  {journal} {\bibinfo  {journal} {Phys.
  Rev. B}\ }\textbf {\bibinfo {volume} {73}},\ \bibinfo {pages} {184505}
  (\bibinfo {year} {2006})}\BibitemShut {NoStop}%
\bibitem [{\citenamefont {Semenov}\ \emph {et~al.}(2016)\citenamefont
  {Semenov}, \citenamefont {Devyatov}, \citenamefont {de~Visser},\ and\
  \citenamefont {Klapwijk}}]{semenov2016}%
  \BibitemOpen
  \bibfield  {author} {\bibinfo {author} {\bibfnamefont {A.~V.}\ \bibnamefont
  {Semenov}}, \bibinfo {author} {\bibfnamefont {I.~A.}\ \bibnamefont
  {Devyatov}}, \bibinfo {author} {\bibfnamefont {P.~J.}\ \bibnamefont
  {de~Visser}}, \ and\ \bibinfo {author} {\bibfnamefont {T.~M.}\ \bibnamefont
  {Klapwijk}},\ }\href {\doibase 10.1103/PhysRevLett.117.047002} {\bibfield
  {journal} {\bibinfo  {journal} {Phys. Rev. Lett.}\ }\textbf {\bibinfo
  {volume} {117}},\ \bibinfo {pages} {047002} (\bibinfo {year}
  {2016})}\BibitemShut {NoStop}%
\bibitem [{\citenamefont {Linder}\ \emph {et~al.}(2016)\citenamefont {Linder},
  \citenamefont {Amundsen},\ and\ \citenamefont {Ouassou}}]{linder2016}%
  \BibitemOpen
  \bibfield  {author} {\bibinfo {author} {\bibfnamefont {J.}~\bibnamefont
  {Linder}}, \bibinfo {author} {\bibfnamefont {M.}~\bibnamefont {Amundsen}}, \
  and\ \bibinfo {author} {\bibfnamefont {J.~A.}\ \bibnamefont {Ouassou}},\
  }\href {\doibase 10.1038/srep38739} {\bibfield  {journal} {\bibinfo
  {journal} {Sci. Rep.}\ }\textbf {\bibinfo {volume} {6}},\ \bibinfo {pages}
  {38739} (\bibinfo {year} {2016})}\BibitemShut {NoStop}%
\bibitem [{\citenamefont {Watts-Tobin}\ \emph {et~al.}(1981)\citenamefont
  {Watts-Tobin}, \citenamefont {Krähenbühl},\ and\ \citenamefont
  {Kramer}}]{tobin1981}%
  \BibitemOpen
  \bibfield  {author} {\bibinfo {author} {\bibfnamefont {R.}~\bibnamefont
  {Watts-Tobin}}, \bibinfo {author} {\bibfnamefont {Y.}~\bibnamefont
  {Krähenbühl}}, \ and\ \bibinfo {author} {\bibfnamefont {L.}~\bibnamefont
  {Kramer}},\ }\href {\doibase 10.1007/BF00117427} {\bibfield  {journal}
  {\bibinfo  {journal} {J. Low. Temp. Phys.}\ }\textbf {\bibinfo {volume}
  {42}},\ \bibinfo {pages} {459} (\bibinfo {year} {1981})}\BibitemShut
  {NoStop}%
\bibitem [{\citenamefont {Kubo}\ and\ \citenamefont
  {Gurevich}(2019)}]{kubo2019}%
  \BibitemOpen
  \bibfield  {author} {\bibinfo {author} {\bibfnamefont {T.}~\bibnamefont
  {Kubo}}\ and\ \bibinfo {author} {\bibfnamefont {A.}~\bibnamefont
  {Gurevich}},\ }\href {\doibase 10.1103/PhysRevB.100.064522} {\bibfield
  {journal} {\bibinfo  {journal} {Phys. Rev. B}\ }\textbf {\bibinfo {volume}
  {100}},\ \bibinfo {pages} {064522} (\bibinfo {year} {2019})}\BibitemShut
  {NoStop}%
\bibitem [{\citenamefont {Buzdin}(2005)}]{buzdin2005}%
  \BibitemOpen
  \bibfield  {author} {\bibinfo {author} {\bibfnamefont {A.~I.}\ \bibnamefont
  {Buzdin}},\ }\href {\doibase 10.1103/RevModPhys.77.935} {\bibfield  {journal}
  {\bibinfo  {journal} {Rev. Mod. Phys.}\ }\textbf {\bibinfo {volume} {77}},\
  \bibinfo {pages} {935} (\bibinfo {year} {2005})}\BibitemShut {NoStop}%
\bibitem [{\citenamefont {Usadel}(1970)}]{usadel1970}%
  \BibitemOpen
  \bibfield  {author} {\bibinfo {author} {\bibfnamefont {K.~D.}\ \bibnamefont
  {Usadel}},\ }\href {\doibase 10.1103/PhysRevLett.25.507} {\bibfield
  {journal} {\bibinfo  {journal} {Phys. Rev. Lett.}\ }\textbf {\bibinfo
  {volume} {25}},\ \bibinfo {pages} {507} (\bibinfo {year} {1970})}\BibitemShut
  {NoStop}%
\bibitem [{\citenamefont {Rammer}\ and\ \citenamefont
  {Smith}(1986)}]{rammer1986}%
  \BibitemOpen
  \bibfield  {author} {\bibinfo {author} {\bibfnamefont {J.}~\bibnamefont
  {Rammer}}\ and\ \bibinfo {author} {\bibfnamefont {H.}~\bibnamefont {Smith}},\
  }\href {\doibase 10.1103/RevModPhys.58.323} {\bibfield  {journal} {\bibinfo
  {journal} {Rev. Mod. Phys.}\ }\textbf {\bibinfo {volume} {58}},\ \bibinfo
  {pages} {323} (\bibinfo {year} {1986})}\BibitemShut {NoStop}%
\bibitem [{\citenamefont {Bergeret}\ and\ \citenamefont
  {Tokatly}(2013)}]{bergeret2013}%
  \BibitemOpen
  \bibfield  {author} {\bibinfo {author} {\bibfnamefont {F.~S.}\ \bibnamefont
  {Bergeret}}\ and\ \bibinfo {author} {\bibfnamefont {I.~V.}\ \bibnamefont
  {Tokatly}},\ }\href {\doibase 10.1103/PhysRevLett.110.117003} {\bibfield
  {journal} {\bibinfo  {journal} {Phys. Rev. Lett.}\ }\textbf {\bibinfo
  {volume} {110}},\ \bibinfo {pages} {117003} (\bibinfo {year}
  {2013})}\BibitemShut {NoStop}%
\bibitem [{\citenamefont {Amundsen}\ and\ \citenamefont
  {Linder}(2017)}]{amundsen2017}%
  \BibitemOpen
  \bibfield  {author} {\bibinfo {author} {\bibfnamefont {M.}~\bibnamefont
  {Amundsen}}\ and\ \bibinfo {author} {\bibfnamefont {J.}~\bibnamefont
  {Linder}},\ }\href {\doibase 10.1103/PhysRevB.96.064508} {\bibfield
  {journal} {\bibinfo  {journal} {Phys. Rev. B}\ }\textbf {\bibinfo {volume}
  {96}},\ \bibinfo {pages} {064508} (\bibinfo {year} {2017})}\BibitemShut
  {NoStop}%
\bibitem [{\citenamefont {Kupriyanov}\ and\ \citenamefont
  {Lukichev}(1988)}]{KL1988}%
  \BibitemOpen
  \bibfield  {author} {\bibinfo {author} {\bibfnamefont {M.~Y.}\ \bibnamefont
  {Kupriyanov}}\ and\ \bibinfo {author} {\bibfnamefont {V.~F.}\ \bibnamefont
  {Lukichev}},\ }\href@noop {} {\ \textbf {\bibinfo {volume} {94}} (\bibinfo
  {year} {1988})}\BibitemShut {NoStop}%
\bibitem [{\citenamefont {Eschrig}\ \emph {et~al.}(2015)\citenamefont
  {Eschrig}, \citenamefont {Cottet}, \citenamefont {Belzig},\ and\
  \citenamefont {Linder}}]{eschrig2015}%
  \BibitemOpen
  \bibfield  {author} {\bibinfo {author} {\bibfnamefont {M.}~\bibnamefont
  {Eschrig}}, \bibinfo {author} {\bibfnamefont {A.}~\bibnamefont {Cottet}},
  \bibinfo {author} {\bibfnamefont {W.}~\bibnamefont {Belzig}}, \ and\ \bibinfo
  {author} {\bibfnamefont {J.}~\bibnamefont {Linder}},\ }\href {\doibase
  10.1088/1367-2630/17/8/083037} {\bibfield  {journal} {\bibinfo  {journal}
  {New J. Phys.}\ }\textbf {\bibinfo {volume} {17}},\ \bibinfo {pages} {083037}
  (\bibinfo {year} {2015})}\BibitemShut {NoStop}%
\bibitem [{\citenamefont {Schmid}\ and\ \citenamefont
  {Schön}(1975)}]{schmid1975}%
  \BibitemOpen
  \bibfield  {author} {\bibinfo {author} {\bibfnamefont {A.}~\bibnamefont
  {Schmid}}\ and\ \bibinfo {author} {\bibfnamefont {G.}~\bibnamefont
  {Schön}},\ }\href {\doibase 10.1007/BF00115264} {\bibfield  {journal}
  {\bibinfo  {journal} {J. Low. Temp. Phys.}\ }\textbf {\bibinfo {volume}
  {20}},\ \bibinfo {pages} {207} (\bibinfo {year} {1975})}\BibitemShut
  {NoStop}%
\bibitem [{\citenamefont {Linder}\ \emph {et~al.}(2008)\citenamefont {Linder},
  \citenamefont {Yokoyama},\ and\ \citenamefont {Sudb\o{}}}]{linder2008}%
  \BibitemOpen
  \bibfield  {author} {\bibinfo {author} {\bibfnamefont {J.}~\bibnamefont
  {Linder}}, \bibinfo {author} {\bibfnamefont {T.}~\bibnamefont {Yokoyama}}, \
  and\ \bibinfo {author} {\bibfnamefont {A.}~\bibnamefont {Sudb\o{}}},\ }\href
  {\doibase 10.1103/PhysRevB.77.174514} {\bibfield  {journal} {\bibinfo
  {journal} {Phys. Rev. B}\ }\textbf {\bibinfo {volume} {77}},\ \bibinfo
  {pages} {174514} (\bibinfo {year} {2008})}\BibitemShut {NoStop}%
\bibitem [{\citenamefont {Blonder}\ \emph {et~al.}(1982)\citenamefont
  {Blonder}, \citenamefont {Tinkham},\ and\ \citenamefont
  {Klapwijk}}]{BTK1982}%
  \BibitemOpen
  \bibfield  {author} {\bibinfo {author} {\bibfnamefont {G.~E.}\ \bibnamefont
  {Blonder}}, \bibinfo {author} {\bibfnamefont {M.}~\bibnamefont {Tinkham}}, \
  and\ \bibinfo {author} {\bibfnamefont {T.~M.}\ \bibnamefont {Klapwijk}},\
  }\href {\doibase 10.1103/PhysRevB.25.4515} {\bibfield  {journal} {\bibinfo
  {journal} {Phys. Rev. B}\ }\textbf {\bibinfo {volume} {25}},\ \bibinfo
  {pages} {4515} (\bibinfo {year} {1982})}\BibitemShut {NoStop}%
\bibitem [{\citenamefont {Asano}\ \emph {et~al.}(2004)\citenamefont {Asano},
  \citenamefont {Tanaka},\ and\ \citenamefont {Kashiwaya}}]{asano2004}%
  \BibitemOpen
  \bibfield  {author} {\bibinfo {author} {\bibfnamefont {Y.}~\bibnamefont
  {Asano}}, \bibinfo {author} {\bibfnamefont {Y.}~\bibnamefont {Tanaka}}, \
  and\ \bibinfo {author} {\bibfnamefont {S.}~\bibnamefont {Kashiwaya}},\ }\href
  {\doibase 10.1103/PhysRevB.69.134501} {\bibfield  {journal} {\bibinfo
  {journal} {Phys. Rev. B}\ }\textbf {\bibinfo {volume} {69}},\ \bibinfo
  {pages} {134501} (\bibinfo {year} {2004})}\BibitemShut {NoStop}%
\bibitem [{\citenamefont {Feshchenko}\ \emph {et~al.}(2015)\citenamefont
  {Feshchenko}, \citenamefont {Casparis}, \citenamefont {Khaymovich},
  \citenamefont {Maradan}, \citenamefont {Saira}, \citenamefont {Palma},
  \citenamefont {Meschke}, \citenamefont {Pekola},\ and\ \citenamefont
  {Zumb\"uhl}}]{feschenko2015}%
  \BibitemOpen
  \bibfield  {author} {\bibinfo {author} {\bibfnamefont {A.~V.}\ \bibnamefont
  {Feshchenko}}, \bibinfo {author} {\bibfnamefont {L.}~\bibnamefont
  {Casparis}}, \bibinfo {author} {\bibfnamefont {I.~M.}\ \bibnamefont
  {Khaymovich}}, \bibinfo {author} {\bibfnamefont {D.}~\bibnamefont {Maradan}},
  \bibinfo {author} {\bibfnamefont {O.-P.}\ \bibnamefont {Saira}}, \bibinfo
  {author} {\bibfnamefont {M.}~\bibnamefont {Palma}}, \bibinfo {author}
  {\bibfnamefont {M.}~\bibnamefont {Meschke}}, \bibinfo {author} {\bibfnamefont
  {J.~P.}\ \bibnamefont {Pekola}}, \ and\ \bibinfo {author} {\bibfnamefont
  {D.~M.}\ \bibnamefont {Zumb\"uhl}},\ }\href {\doibase
  10.1103/PhysRevApplied.4.034001} {\bibfield  {journal} {\bibinfo  {journal}
  {Phys. Rev. Applied}\ }\textbf {\bibinfo {volume} {4}},\ \bibinfo {pages}
  {034001} (\bibinfo {year} {2015})}\BibitemShut {NoStop}%
\bibitem [{\citenamefont {Li}\ \emph {et~al.}(2019)\citenamefont {Li},
  \citenamefont {de~Ronde}, \citenamefont {de~Boer}, \citenamefont {Ridderbos},
  \citenamefont {Zwanenburg}, \citenamefont {Huang}, \citenamefont {Golubov},\
  and\ \citenamefont {Brinkman}}]{li2019}%
  \BibitemOpen
  \bibfield  {author} {\bibinfo {author} {\bibfnamefont {C.}~\bibnamefont
  {Li}}, \bibinfo {author} {\bibfnamefont {B.}~\bibnamefont {de~Ronde}},
  \bibinfo {author} {\bibfnamefont {J.}~\bibnamefont {de~Boer}}, \bibinfo
  {author} {\bibfnamefont {J.}~\bibnamefont {Ridderbos}}, \bibinfo {author}
  {\bibfnamefont {F.}~\bibnamefont {Zwanenburg}}, \bibinfo {author}
  {\bibfnamefont {Y.}~\bibnamefont {Huang}}, \bibinfo {author} {\bibfnamefont
  {A.}~\bibnamefont {Golubov}}, \ and\ \bibinfo {author} {\bibfnamefont
  {A.}~\bibnamefont {Brinkman}},\ }\href {\doibase
  10.1103/PhysRevLett.123.026802} {\bibfield  {journal} {\bibinfo  {journal}
  {Phys. Rev. Lett.}\ }\textbf {\bibinfo {volume} {123}},\ \bibinfo {pages}
  {026802} (\bibinfo {year} {2019})}\BibitemShut {NoStop}%
\end{thebibliography}%


%

\end{document}


\title{Supplementary: Temporarily enhanced superconductivity from magnetic fields}

\author{Eirik Holm Fyhn}
\affiliation{Center for Quantum Spintronics, Department of Physics, Norwegian \\ University of Science and Technology, NO-7491 Trondheim, Norway}
\author{Jacob Linder}
\affiliation{Center for Quantum Spintronics, Department of Physics, Norwegian \\ University of Science and Technology, NO-7491 Trondheim, Norway}

\date{\today}
\maketitle
\section{Kinetic equations}%
\label{sec:kinetic_equations}
Finding the retarded Green's function is enough to calculate the local density of states, but for many other quantities, such as charge or spin currents, magnetization or Cooper-pair correlation functions, one needs the full Keldysh Green's function.
Here we show the equations for the distribution function $h$, which can be used to find the Keldysh Green's function through the relations
\begin{align}
\label{eq:rels}
  \hat g^A = -\hat \rho_3\left(\hat g^R\right)^\dagger\hat \rho_3,
  \qquad
  \hat g^K = \hat g^R\circ h - h\circ \hat g^A.
\end{align}

We start by writing $h$ as an expansion in the small parameter, 
\begin{equation}
  h = \sum_{n=0}^\infty z^n h_n.
\end{equation}
In order to solve for the charge current in a way that consistently include the supercurrent contribution we need to solve for $h$ to second order in $z$.
This can be seen from the fact that charge and spin currents are given by the diagonal components of
\begin{multline}
  \label{eq:currKeldysh}
  \hat j^K = \hat g^R\circ \left(\tilde \nabla\circ \hat g^R\right)\circ h - h\circ \hat g^A\circ \left(\tilde \nabla\circ \hat g^A\right)\\ 
  + \tilde\nabla\circ h - \hat g^R\circ \left(\tilde\nabla\circ h\right)\circ \hat g^A,
\end{multline}
where $\hat j^K$ is the upper right block of $\check j \defeq \check g \circ \tilde\nabla\circ \check g$.
The supercurrent, which are given by the first two terms on the right hand side of \cref{eq:currKeldysh}, is at least second order in $z$.
Hence, one must in general find $h_0$, $h_1$ and $h_2$.

The reason why the perturbation expansion in $z$ works for $\hat g^R$, is that $\hat g^R$ is independent of $\varepsilon$ to zeroth order.
This is not true for $h$.
However, the equation for $h$ is linear, so the perturbation expansion is nevertheless able to remove circle-products between functions that depend on $\varepsilon$.
The equation for $h$ can be found by taking the covariant derivative of \cref{eq:currKeldysh} and inserting the Usadel equation presented in the main text.
We find that
\begin{multline}
  \label{eq:hEnergy}
  \hat\rho_3\frac{\partial h}{\partial T}\circ\hat g^A - \hat g^R \circ \frac{\partial h}{\partial T}\hat\rho_3
  + \tilde\nabla\circ\tilde\nabla\circ h + \hat j^R\circ\tilde\nabla\circ h 
 \\
  - \left(\tilde\nabla\circ h\right)\circ \hat j^A
  - \hat g^R\circ \left[\tilde\nabla\circ\tilde\nabla\circ h\right] \circ \hat g^A \\
  - \left(\tilde\nabla\circ \hat g^R\right)\circ \left(\tilde\nabla\circ h\right)\circ \hat g^A 
  - \hat g^R\circ \left(\tilde\nabla\circ h\right)\circ \tilde\nabla\circ \hat g^A \\
  = i\comm{\check \sigma}{\check g}_\circ^R \circ h - ih\circ \comm{\check \sigma}{\check g}_\circ^A - i\comm{\check \sigma}{\check g}_\circ^K,
\end{multline}
where $\hat j^R$ and $\hat j^A$ are the upper left and lower right blocks of $\check j$ and $\varepsilon\diag(\hat \rho_3, \hat\rho_3)$ has been extracted from the self-energy $\check\sigma$.
The commutators are with respect to the circle-product and can be evaluated using
\begin{multline}
  i\comm{\check \sigma}{\check g}_\circ^R \circ h - ih\circ \comm{\check \sigma}{\check g}_\circ^A - i\comm{\check \sigma}{\check g}_\circ^K \\
  =i\hat g^R\circ\left[\hat \sigma^K - \left(\hat \sigma^R\circ h - h\circ \hat \sigma^A\right)\right] \\
  - i\left[\hat \sigma^K - \left(\hat \sigma^R\circ h - h\circ \hat \sigma^A\right)\right]\circ \hat g^A.
\end{multline}
We will assume that $\hat\sigma^K$ is block-diagonal, just like $\hat\sigma^R$ and $\hat\sigma^A$.
This assumption is valid in the system considered in the main manuscript.
Unlike $\hat\sigma^R$ and $\hat\sigma^A$, however, there will be no restrictions on $\hat\sigma^K$ with regards to its energy-dependence.
This is taken advantage of in the relaxation time approximation used in the main text.
Note that the same derivation can be done when $\hat\sigma^K$, $\hat\sigma^R$ and $\hat\sigma^A$ are not block-diagonal.
The only difference is which terms to include in the perturbation expansion of \cref{eq:hEnergy}.

The Keldysh part of Kupriyanov-Lukichev boundary condition can be written
\begin{widetext}
\begin{align}
  \uv n \vdot
  \left[\tilde\nabla\circ h - \hat g^R\circ\left(\tilde\nabla\circ h\right)\circ \hat g^A\right] = \frac{z}{2}
  \Biggl\{\hat g^R\circ\left[\hat g_s^R\circ(h_s-h)-(h_s-h)\circ \hat g_s^A\right] 
  -\left[\hat g_s^R\circ(h_s-h)-(h_s-h)\circ \hat g_s^A\right]\circ \hat g^A\Biggr\},
\end{align}
\end{widetext}
where $h_s$ is the distribution function in the neighbouring region.

We find that to zeroth order in $z$,
\begin{subequations}
  \label{eq:h0_energy}
  \begin{align}  
    \frac{\partial h_0}{\partial T} = \tilde\nabla\circ\tilde\nabla\circ h_0
    - i\hat\rho_3 \left(\hat\sigma^K - \hat\sigma^R\circ h_0 + h_0\circ \hat\sigma^A\right), \\ 
    \uv n \cdot \tilde\nabla \circ h_0 = 0,
  \end{align}  
\end{subequations}
and to first order,
\begin{subequations}
  \label{eq:h1_energy}
  \begin{align}  
    \frac{\partial h_1}{\partial T} = \tilde\nabla\circ\tilde\nabla\circ h_1
    + i\hat\rho_3 \left(\hat\sigma^R\circ h_1 - h_1\circ \hat\sigma^A\right), \\ 
    \uv n \cdot \tilde\nabla \circ h_1 = \frac 1 4 \Bigl\{ \hat\rho_3 \left[\hat g_s^R\circ(h_s-h_0)-(h_s-h_0)\circ \hat g_s^A\right] \nonumber\\ 
    +\left[\hat g_s^R\circ(h_s-h_0)-(h_s-h_0)\circ \hat g_s^A\right]\hat\rho_3\Bigr\}.
    \label{eq:h1_bc_en}
  \end{align}  
\end{subequations}
There is also a first order equation of off-diagonal matrices, but this is automatically satisfied from \cref{eq:h0_energy}.
Finally, $h_2$ satisfies
\begin{subequations}
  \label{eq:h2_energy}
  \begin{align}  
    2\frac{\partial h_2}{\partial T} = 2\tilde\nabla\circ\tilde\nabla\circ h_2
    + 2i\hat\rho_3 \left(\hat\sigma^R\circ h_2 - h_2\circ \hat\sigma^A\right) \nonumber\\
    + \hat j^R_2 \circ \left(\tilde\nabla\circ h_0\right) 
    - \left(\tilde\nabla\circ h_0\right)\circ\hat j_2^A
    + \left(\tilde\nabla\circ\hat g_2\right) \circ\tilde\nabla\circ h_0 \hat\rho_3\nonumber
    \\
    + \hat\rho_3\left(\tilde\nabla\circ h_0\right)\circ \tilde\nabla\circ \hat g_2^\dagger
    - \left(\tilde\nabla\circ \hat f_1\right) \circ\left(\tilde\nabla\circ h_0\right) \circ \hat f_1^\dagger
    \nonumber \\
    - \hat f_1 \circ\left(\tilde\nabla\circ h_0\right) \circ \tilde\nabla\circ \hat f_1^\dagger, \\
    \uv n \cdot \tilde\nabla \circ h_2 = \frac 1 4 \Bigl\{ \hat f_1\circ \left[\hat g_s^R\circ(h_s-h_0)-(h_s-h_0)\circ \hat g_s^A\right] \nonumber \\ 
    -\left[\hat g_s^R\circ(h_s-h_0)-(h_s-h_0)\circ \hat g_s^A\right]\circ \hat f_1^\dagger \nonumber \\
    -\hat\rho_3 \left[\hat g_s^R\circ h_1-h_1\circ \hat g_s^A\right]  
    -\left[\hat g_s^R\circ h_1-h_1\circ \hat g_s^A\right]\hat\rho_3
  \Bigr\}.
    \label{eq:h2_bc_en}
  \end{align}  
\end{subequations}
where
\begin{subequations}
\begin{align}
  \hat j_2^R = \hat f_1\circ\tilde\nabla \circ \hat f_1 + \rho_3\tilde\nabla \circ \hat g_2,\\
  \hat j_2^A = \hat f_1^\dagger\circ\tilde\nabla \circ \hat f_1^\dagger + \rho_3\tilde\nabla \circ \hat g_2^\dagger.
\end{align}
\end{subequations}
\Cref{eq:h1_bc_en,eq:h2_bc_en} can be further simplified by noting that, since $h_1$ and $h_2$ are block-diagonal, so too must $\left[\hat g_s^R\circ(h_s-h_n)-(h_s-h_n)\circ \hat g_s^A\right]$ for $n=1$ and $n=2$.
As a result they commute with $\hat\rho_3$.
Additionally,
\begin{align*}
0 = \hat f_1\circ \left[\hat g_s^R\circ(h_s-h_0)-(h_s-h_0)\circ \hat g_s^A\right] \\ 
    -\left[\hat g_s^R\circ(h_s-h_0)-(h_s-h_0)\circ \hat g_s^A\right]\circ \hat f_1^\dagger
\end{align*}
because $h_2$ would have off-diagonal components otherwise.
Hence,
\begin{align}
  \uv n \cdot \tilde\nabla \circ h_1 &= \frac{\hat\rho_3}{2} \left[\hat g_s^R\circ(h_s-h_0)-(h_s-h_0)\circ \hat g_s^A\right], \\
  \uv n \cdot \tilde\nabla \circ h_2 &= -\frac{\hat\rho_3}{2} \left[\hat g_s^R\circ h_1-h_1\circ \hat g_s^A\right].
  \label{eq:bc:upd}
\end{align}

Unlike the equation for $\hat f_1$, presented in the main text, the equations for the distribution functions has circle-products that do not reduce to ordinary matrix products in Fourier space.
However, these are all circle-products of functions that can be evaluated prior to solving the equations.
This suggests an order in which to solve the equations.
One can find $\hat f_1$ and $h_0$ first, but in order to find $h_1$ one must first know $h_0$ and in order to find $h_2$ one must have solved $\hat f_1$, $h_0$ and $h_1$.
To write \cref{eq:h0_energy,eq:h1_energy,eq:h2_energy} in a way that does not require evaluation of circle-products between unknown functions is now only a matter of Fourier-transforming, writing at the covariant derivatives and using eq.~(11) in the main text.

Note that the equation for $\hat f_1$ involve differentiation with respect to $t$, whereas the center of mass time $T$ appear only as a parameter. The equations for the distribution functions $h_0$, $h_1$ and $h_2$ are opposite in this regard, and involve differentiation with respect to $T$ but not $t$.

\section{Analytical solution to eq. (14)}%
\label{sec:analytical_solution_to_eq_16_}
The retarded and Keldysh self-energies for the SNS-junction with inelastic scattering and time-dependent and spatially homogeneous exchange field $m(T)$ are
\begin{subequations}
  \begin{align}
    \hat\sigma^R &= i\delta\hat\rho_3 + m \diag(1,-1,1,-1),\\
    \hat\sigma^K &= 2i\delta\hat\rho_3 h_\text{eq},
  \end{align} 
\end{subequations}
where $\delta$ is the inelastic scattering rate, $h_\text{eq}(\varepsilon) = \tanh(\beta\varepsilon/2)$ and $\beta$ is the inverse temperature towards which the system relaxes.

From \cref{eq:h0_energy,eq:h1_energy,eq:h2_energy} we see that $h_0$, $h_1$ and $h_2$ only have non-zero components proportional to the identity matrix $I_4$ and $\diag(1,-1,-1,1)$.
Hence, only the supercurrent contributes to the charge current $\propto \int\dd{\varepsilon} \Tr(\hat\rho_3\hat j^K)$.
For this reason we need only find the retarded Green's function and $h_0$.
We repeat the relevant equations here for convenience.
If we write the upper right block of $\hat F_1$ as $F_1 = \sigma_1F_t + \sigma_2 F_s$, where $\sigma_1$ and $\sigma_2$ are Pauli matrices, the zeroth order distribution function $H_0 = H_L I_4 + H_{TS}\diag(1,-1,-1,1)$ and let $m^{\pm}(t,T) = m(T+t/2) \pm m(T-t/2)$, we find that
\begin{subequations}
  \label{eq:exchange}
  \begin{align}
  \left(2\frac{\partial}{\partial t} - \nabla^2 + 2\delta\right)\mqty(F_s \\ F_t)
  = \mqty(-m^+ F_t \\ m^+ F_s), \\
  \left.\uv n \cdot \nabla F_s \right\rvert_{x=0,1} = F^\text{BSC}_{l,r}, \quad 
  \left.\uv n \cdot \nabla F_t \right\rvert_{x=0,1} = 0 \\
  \left(\frac{\partial}{\partial T} + 2\delta\right)\mqty(H_L-H_\text{eq} \\ H_{TS})
  = \mqty(-m^- H_{TS} \\ m^- H_L), 
  \end{align}
\end{subequations}
where $F^\text{BCS}_l = \Delta\me{-\delta t} J_0(\abs{\Delta}t)\theta(t)$ and $F^\text{BCS}_r = \me{i\phi}F^\text{BCS}_l$ are the anamalous Green's functions in the left and right superconductors, respectively. $J_0$ is the zeroth order Bessel function of the first kind.
The superconducting energy gap is $\Delta$, and $\phi$ is the phase difference between the two superconductors.
Finally, with the notation used in the main text $m^\pm(t,T) = m(T+t/2)\pm m(T-t/2)$.

Assuming that the system is at equilibrium at $T = -\infty$, we find that the solution is
\begin{subequations}
  \label{eq:sol}
  \begin{align}  
    F_s = \frac{\Delta\me{-\delta t}}{2}\int_0^t \dd{\tau} J_0(\abs{\Delta}\tau)
  \cos(\frac 1 2 \int_{\tau}^t\dd{\tilde\tau}m^+(\tilde\tau, T)) \nonumber\\ 
  \times \sum_{n=-\infty}^\infty\left[1+(-1)^n\me{i\phi}\right] \cos(n\pi x)\me{-\frac 1 2 n^2\pi^2(t-\tau)}, \\
    F_t = \frac{\Delta\me{-\delta t}}{2}\int_0^t \dd{\tau} J_0(\abs{\Delta}\tau)
  \sin(\frac 1 2 \int_{\tau}^t\dd{\tilde\tau}m^+(\tilde\tau, T)) \nonumber\\ 
  \times \sum_{n=-\infty}^\infty\left[1+(-1)^n\me{i\phi}\right] \cos(n\pi x)\me{-\frac 1 2 n^2\pi^2(t-\tau)}, \\
  H_L = 2\delta H_\text{eq} \int_{-\infty}^T\dd{\tau} \me{-2\delta (T-\tau)}
  \cos(\int_\tau^T\dd{\tilde\tau} m^-(t,\tilde\tau)),\\
  H_{TS} = 2\delta H_\text{eq} \int_{-\infty}^T\dd{\tau} \me{-2\delta (T-\tau)}
  \sin(\int_\tau^T\dd{\tilde\tau} m^-(t,\tilde\tau)).
  \end{align}  
\end{subequations}

\section{Observables}%
\label{sec:observables}
The singlet Cooper pair correlation function $\Psi_s$ and electrical current $I$ can be obtained from inserting the analytical solution given by \cref{eq:sol} into the expressions
\begin{equation}
  \Psi_s  = \left.-i\pi N_0\left(F_s\bullet H_L - F_t\bullet H_{TS}\right)\right\rvert_{t=0},
\end{equation}
and
\begin{equation}
  I = \frac{\pi e D}{4}\Tr\left[\hat\rho_3\left(\check G \bullet \tilde\nabla\bullet \check G\right)^K\right]_{t=0}
  ,
\end{equation}
After some algebra we find that
\begin{align}
  I = \pi N_0 eD\Bigl[\Im\bigl\{
    & F_s\bullet \nabla F_s^* + \nabla F_s^*\bullet F_s
    \nonumber\\
  -&F_t\bullet \nabla F_t^* - \nabla F_t^*\bullet F_t\bigr\}\bullet iH_L 
  \nonumber\\
  -\Im\bigl\{&F_t\bullet \nabla F_s^* + \nabla F_s^*\bullet F_t
      \nonumber\\
  + & F_s\bullet \nabla F_t^* + \nabla F_t^*\bullet F_s\bigr\}\bullet iH_{TS}
\Bigr]_{t=0}.
\end{align}

To understand the non-equilibrium behaviour it is useful to use the distribution functions for spin-up, $H_+ = H_L + iH_{TS}$, and spin-down, $H_- = H_L - iH_{TS}$.
If we write the upper right block of $\hat F = \antidiag(F_+, F_-)$, then $F_\pm = F_t \mp i F_s$.
With this we get that the singlet Cooper pair correlation function can be written
\begin{equation}
  \Psi = \frac{\pi N_0}{2}\left(F_+ \bullet H_+ - F_- \bullet H_-\right)\lvert_{t=0},
\end{equation}
and the current is
\begin{equation}
  I = \frac{\pi N_0 e D}{4} \left(J_+\bullet H_+ + J_-\bullet H_-\right)\rvert_{t=0},
\end{equation}
where
\begin{subequations}
  \label{eq:specCurr}
  \begin{align}
    J_+ = F_-^* \bullet\nabla F_+ + \nabla F_+\bullet F_-^* - F_+\bullet\nabla F_-^* - \nabla F_-^*\bullet F_+, \\
    J_- = F_+^* \bullet\nabla F_- + \nabla F_-\bullet F_+^* - F_-\bullet\nabla F_+^* - \nabla F_+^*\bullet F_-.
  \end{align}
\end{subequations}
We note in passing that both $J_+$ and $J_-$ are proportional to $\sin\phi$, as can be seen most readily by evaluating them at $x = 1/2$.
Hence, the critical current occurs when $\phi$ is a half-integral multiple of $\pi$.
Also, when written as function of the exchange field $m$ we have that $F_-(-m) = -F_+(m)$ and $J_-(-m) = J_+(m)$.

In order to study how the system evolves immediately after the exchange field is turned on, we can set the inelastic scattering rate to 0.
In this case we find that 
\begin{equation}
  H_\pm(t, T) = H_\text{eq}(t) \exp(\pm i \int_{T-t/2}^{T+t/2}\dd{\tau} m(\tau)).
\end{equation}
Assuming that $T > \abs{t}/2$ for all the relevant relative times $t$, this is simply $H_\pm = H_\text{eq}\me{\pm i m_0 t}$.
Hence, in energy space we have $h_\pm(\varepsilon) = h_\text{eq}(\varepsilon \pm m_0) = \tanh[\beta(\varepsilon \pm m_0)/2]$.

If $T > \abs{t}/2$ for all the relevant relative times $t$ we can also take advantage of the fact that the system is stationary, such that the circle-products in energy space reduces to normal multiplications.
That is,
\begin{equation}
  \Psi = \frac{N_0}{4} \int_{-\infty}^{\infty} \dd{\varepsilon} (f_+ h_+ - f_- h_-),
\end{equation}
and
\begin{equation}
  I = \frac{N_0 e D}{8} \int_{-\infty}^{\infty} \dd{\varepsilon} (j_+ h_+ + j_- h_-).
\end{equation}

\begin{figure}[]
  \centering
  \includegraphics[width=1.0\linewidth]{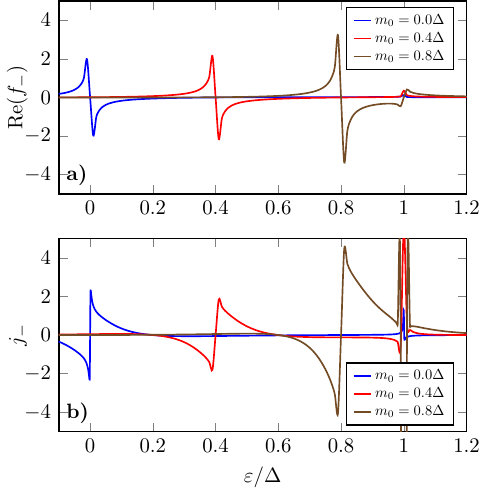}
  \caption{a) The real part of the anomalous Green's function $f_-$ with $\phi =0$ and b) spectral current $j_-$ with $\phi=\pi/2$ as a function of energy $\varepsilon$ for various exchange fields $m_0$. The energy gap in the superconductors is $\Delta = 50\thouless$, where $\thouless$ is the Thouless energy, and the inelastic scattering rate is $\delta=\Delta\times 10^{-3}$.}%
  \label{fig:integrands}
\end{figure}

From \cref{eq:specCurr} we can immediately see that the spectral current is antisymmetric in energy when $m_0=0$.
In this case $J_+=J_-=-4i\Im(F_s^*\bullet\nabla F_s)$ is purely imaginary, so the real part of its Fourier transform must be antisymmetric in $\varepsilon$.
Because only the real part of the spectral current can give rise to a real electrical current, the spectral current is antisymmetric around $\varepsilon=0$ when $m_0=0$.
Similarly, $F_\pm$ are purely imaginary when $\phi=0$, $m_0=0$ and $\Delta$ is real, so the same arguments shows that the real parts of $f_\pm$ are antisymmetric around $\varepsilon = 0$.
It is the real part of $f_\pm$ that is relevant for $\Psi_s$ when $\phi=0$, since $\Psi_s$ is real.

When $m_0 \neq 0$ and $T > \abs{t}/2$ we get that
\begin{align}
  F_\pm =& \mp\frac{i\Delta\me{-\delta t}\me{\pm im_0 t}}{2}\int_0^t \dd{\tau} J_0(\abs{\Delta}\tau)
  \me{\pm im_0\tau} \nonumber\\ 
         &\times \sum_{n=-\infty}^\infty\left[1+(-1)^n\me{i\phi}\right] \cos(n\pi x)\me{-\frac 1 2 n^2\pi^2(t-\tau)}.
\end{align}
The factor $\me{\pm im_0 t}$ gives rise to a shift $\varepsilon \to \varepsilon \pm m$ in the Fourier transform.
Hence, $f_\pm(\varepsilon) = p_\pm(\varepsilon \pm m_0 + i\delta)$, where $p_\pm$ is the Fourier transform of the remaining integral.
When $m_0 \ll \Delta$, the presence of $\me{\pm i m_0 \tau}$ in the integrand works to shift oscillations with frequency $\Delta$, but this does not affect the low frequency components.
To see why, note that $J_0(x)$ is well approximated by $\cos(x-\pi/4)\sqrt{2/\pi x}$ for large $x$ and
\begin{equation}
  \cos(a)\me{ib} = \frac{\me{i(a+b)}+\me{-i(a-b)}}{2}.
\end{equation}
Hence, the oscillatory part of the integrand is shifted to $\Delta\pm m_0$ while the low-frequency components in $p_\pm$ are left unchanged.
Because we know that $p_\pm$ is antisymmetric for $m_0=0$, we see that $f_\pm$ is antisymmetric close to $\varepsilon \pm m_0$ as long as $m_0 \ll \Delta$.
When  $m_0 \approx \Delta$, the oscillations in $J_0(\abs{\Delta}\tau)$ are matched by those in $\me{\pm i m_0\tau}$, which affects the low-frequency components of $p_\pm$.

\Cref{fig:integrands} shows $\Re(f_-)$ and $j_-$ for various values of $m_0$.
From \cref{fig:integrands} one can see that the integrands are antisymmetric and large close to $\varepsilon = m_0$, which agrees with the discussion above.
Moreover, we see that the values close to $m_0$ are enhanced as $m_0$ approaches $\abs{\Delta}$.
This can be understood as coming from the fact that Andreev reflections with long range occur at the energies where the probability of Andreev reflections is larger, as discussed in the main text.
In equilibrium $h_\pm = \tanh(\beta\varepsilon/2)$, so the positive and negative contributions cancel when $m_0 > 0$.
However, out of equilibrium the distribution function changes sign at $\varepsilon = m_0$, which allows for a significant contribution from the part close to $\varepsilon = m_0$.

\section{Consistency with Maxwell's equations}%
\label{sec:consistency_with_maxwell_s_equations}
Having solved the Usadel equation, the next step is to ensure that the solution is consistent with Maxwell's equation.
There are no orbital effects in the one-dimensional wire, but there could in principle be an induced electric field coming from a nonuniform charge distribution.
The charge density $\rho$ can be calculated from
\begin{equation}
  \rho(T) = -2eN_0\left\{\frac{\pi}{8}\Tr[\hat G^K(0, T)] + 2e\phi(T)\right\},
\end{equation}
where $\phi$ is the electrochemical potential.

Since $H_0$ and $H_1$ has no component proportional to $\hat\rho_3$, we see that $\rho$ is at least second order in $z$.
Thus the induced electric field is also at least second order in $z$ and thereby does not affect $\hat F_1$, $H_0$ or $H_1$.
For this reason it follows that it does not affect the Cooper pair correlation function or the supercurrent.
Nevertheless, it can in principle give rise to a resistive current contribution,
\begin{equation}
  I_r = z^2\frac{\pi e D}{2}\nabla\Tr\left[\hat\rho_3 H_2(0, T)\right].
\end{equation}

By using the fact that the equilibrium distribution function at electrochemical potential $\phi$ is
\begin{equation}
  H_\text{eq}(t,T) = \frac{-i}{\beta\sinh(\pi t/\beta)}\exp(-ie\hat\rho_3 \int_{T-t/2}^{T+t/2}\dd{\tau} \phi\left(\tau\right)),
\end{equation}
we get from \cref{eq:h2_energy} that the resistive current solves
\begin{equation}
  \frac{\partial I_r}{\partial(T\thouless)} - \frac{\partial^2 I_r}{\partial(x/L)^2} 
  = 4z^2 \delta L^2 e^2 E,
\end{equation}
where $x$, $T$ and $\delta$ are now not dimensionless and $E = -\nabla\phi$ is the electric field.
From \cref{eq:bc:upd} we get that the boundary condition is simply $I_r = 0$ at the interfaces.
Thus, this resistive current does not contribute to the total current going through the wire and instead acts to redistribute charge.

By calculating $\Tr[\hat G^K]$ from the analytical expression we find that the $I_r$ is negligible and can be safely ignored.

\section{Beyond the dirty limit}%
\label{sec:beyond_the_dirty_limit}
The focus so far has been on the so-called dirty limit, where the elastic scattering time is small, such that the isotropic part of the Green's function dominates.
However, the same framework as was presented here and in the main text can be applied also for systems outside the dirty limit.
In fact, it works even better in the clean limit.

With arbitrary elastic scattering time $\tau_e$, the quasiclassical Green's function solves the Eilenberger equation,
\begin{equation}
  i\v v_F \cdot \tilde\nabla\circ \check g + \left[\varepsilon\hat\rho_3 + \check\sigma + \frac{i}{2\tau_e} \check g_s, \check g\right]_\circ = 0,
  \label{eq:eilenberger}
\end{equation}
where $\v v_F$ is the Fermi velocity, the subscript $\circ$ denotes that the commutator is taken with respect to the circle product and $\check g_s$ is the isotropic part of the Green's function.
\Cref{eq:eilenberger} gives the Usadel equation in the limit $\tau_e\to 0$.

The method presented in the main text works by writing the equations in the $(t, T)$-coordinates and eliminating circle products between functions that both depend on energy, except the term in the commutator that is linear in $\varepsilon$.
This was done by linearizing the equations in the proximity effect.
We see from \cref{eq:eilenberger} that the same is possible for arbitrary impurity concentrations.
The difference is that in the Usadel equation it was the term $D\tilde\nabla\circ \left(\check g \circ \tilde\nabla\circ \check g\right)$ that required linearization, whereas in the Eilenberger equation it is $i[\check g_s, \check g]/2\tau_e$.
This term is absent in the clean limit since $\tau_e \to \infty$, so in this case \cref{eq:eilenberger} is automatically free from the difficult kind of circle-products if $\check\sigma$ does not depend on energy.
This means that we do not need to linearize in the proximity effect.
One consequence of this is that in the clean limit it could be possible to solve time-dependent transient situations even with retarded self-energies that are not block-diagonal to lowest order in $z$.
In particular, this means that one could add superconducting pairing in the self-energy.
Thus, one could potentially use this framework to find transient phenomena in clean superconductors.

Whether the full equations can be solved in time-dependent situations without additional simplifying assumption depends on the boundary conditions.
One type of boundary condition which can be used in the clean limit is Zaitsev's linearized boundary condition [A. Zaitsev, Zh. Eksp. Teor. Fiz. \textbf{86}, 1742-1758], valid for a weak proximity effect.
In that case, the antisymmetric part of the anomalous Green's function is continuous across the interfaces whereas the symmetric part has a drop proportional to the antisymmetric part of the anomalous Green function [N. Garcia and L.R. Tagirov, arXiv:cond-mat/0601212].
For such a boundary condition, the clean-limit equations can be solved even for an arbitrary time-dependence.